\documentclass[conference]{IEEEtran}
\IEEEoverridecommandlockouts

 \usepackage{cite}

 \usepackage[justification=centering]{caption} 
 
 \usepackage[T1]{fontenc}

\usepackage[caption=false]{subfig}
\usepackage{color}
\usepackage[usenames,dvipsnames,svgnames,table]{xcolor}
\usepackage{multirow}
\usepackage{rotating}
\usepackage{graphicx}
\usepackage[linesnumbered, vlined, ruled]{algorithm2e}
\usepackage{booktabs}
\usepackage{amssymb}
\usepackage{amsfonts}
\usepackage{amsmath}
\usepackage{soul}
\usepackage{etex}
\usepackage{extarrows}
\usepackage{relsize}
\usepackage{epstopdf}

\usepackage{url}
\usepackage[bookmarks=false]{hyperref}
\hypersetup{ 
    colorlinks,%
    citecolor=black,%
    filecolor=black,%
    linkcolor=black,%
    urlcolor=black,%
pdftitle={graphVizdb  A Scalable Platform for Interactive Large Graph Visualization},
pdfauthor={Nikos Bikakis, John Liagouris, Maria Krommyda, George Papastefanatos, Timos Sellis},
pdfsubject={graphVizdb  Scalable Interactive Large Graph and Network Visualization},
pdfkeywords={
big data,  network visualization, massive graphs,  graph drawing,   visual analytics, database visualization, visualization tool, visual exploration,   graph layout, network analysis,  graph summarization, disk based visualization, spatial indexing, large scale graph,  graphvizdb, large graph, R-tree, graph visualization, RDF visualization, network abstraction, Web of Data, data reduction, million nodes graphs, Linked Data visualization, large scale network, scalable visualizations, Semantic Web, visualizing very large datasets, billion nodes graph, interactive data exploration, scalable navigation, RDF graph, multilevel exploration, billion edges, interactive navigation, million edges, multiscale visualization, graph keyword search,  incremental visualization, graph prefetching, scalable visual operations, scalable navigation, efficient exploration, efficient visualization, scalable graph layout, partition-based graph drawing.
 }
}

\newcommand{\eat}[1]{}

\newcommand{\myFontC}[1]{{\fontfamily{pcr}\selectfont #1}} 
\newcommand{\myFontH}[1]{{\fontfamily{phv}\selectfont #1}} 

\newcommand{\tline}  {\specialrule{0.8 pt}{0pt}{1pt}}		 
\newcommand{\bline}  {\specialrule{0.8 pt}{1pt}{0pt}}		
\newcommand{\dline}  {\specialrule{0.4 pt}{0pt}{2pt} \specialrule{0.4 pt}{0pt}{2pt}}		

\definecolor{dark-gray}{gray}{0.3}

\newcommand{\mycomment} [1] 	{\textcolor{dark-gray}{ \tiny {\textsf{#1}} }}
\newcommand{\stitle}[1]{\vspace{0cm}\noindent\textbf{#1}}

\newcounter{def}

\newcounter{lem}

\newcounter{claimcnt}

\newcounter{propcnt}

\newcounter{propertycnt}

\newcounter{ther}

\newcounter{prob}

\newcounter{enum}

\newcounter{myCnt}

\SetKw{Logand}{and}	
\SetKw{Logor}{or}	
\SetKw{Lognot}{not}	

\SetKw{mybreak}{break}	
\SetKw{mycontinue}{continue}	

\SetKwInput{KwVar}{Variables}
\SetKwInput{KwPar}{Parameters}

\SetKwData{myEOF}{EOF}	
\SetKwData{myfalse}{false}
\SetKwData{mytrue}{true}
\SetKwData{mynull}{null}
\SetKwData{myempty}{empty}
\SetKwBlock{myblock}{beginnikos}{end}

\SetAlgoInsideSkip{} 
\SetAlgoSkip{} 
\SetVlineSkip {0.5mm} 
\SetAlgoCaptionSeparator{.}
\SetKwComment{Comment}{\mycomment{\#}}{}
\SetCommentSty{textsf }
\DontPrintSemicolon
\SetAlgoCaptionLayout{small}

\begin{document}

\title{
graphVizdb: A Scalable Platform for Interactive Large Graph Visualization 
 }

\author{
\IEEEauthorblockN{
Nikos  Bikakis$^{\dagger\yen}$ \hspace{22pt} %
John Liagouris$^{\#}$ \hspace{25pt} 
Maria Krommyda${^\dagger}$  \vspace{5pt} \\  
George Papastefanatos$^{\yen}$ \hspace{22pt} 
 Timos Sellis$^{\star}$
} 
\\
%
\IEEEauthorblockA{ 
{$^{\dagger}$NTU  Athens, Greece}
 \hspace{5pt}
{$^{\yen}$ATHENA Research Center, Greece}
\vspace{1pt}
 \\
{$^{\#}$Dept.\ of Computer Science, ETH Z\"urich, Switzerland} \hspace{5pt}
$^{\star}$Swinburne University of Technology, Australia
}
}
 
\maketitle


\begin{abstract}
We present a novel platform for the interactive visualization of very large graphs.
The platform enables the user to interact with the visualized graph in a way that is very similar to the exploration of maps at multiple levels. Our approach involves an offline preprocessing phase that builds the layout of the graph by assigning coordinates to its nodes with respect to a Euclidean plane. The respective points are indexed with a spatial data structure, i.e., an R-tree, and stored in a database.
Multiple abstraction layers of the graph based on various
criteria are also created offline, and they are indexed similarly so
that the user can explore the dataset at different levels of granularity,
depending on her particular needs.
Then, our system translates user operations into simple and very efficient spatial operations (i.e., window queries)   in the backend. This technique allows for a fine-grained access to very large graphs with extremely low latency and memory requirements and without compromising the functionality of the tool.
Our web-based prototype supports three main operations: (1) interactive navigation, (2) multi-level exploration, and (3) keyword search on the graph metadata.


\end{abstract}



\section{Introduction}
\label{sec:intro}

Graph visualization is a core task in various applications such as scientific data management, social network analysis, and decision support systems.
With the wide adoption of the RDF data model  and the recent \emph{Linked Open Data} initiative, graph data are almost everywhere \cite{bikakis15}.
Visualizing these data as graphs provides the non-experts with an intuitive means to explore the content of the data, identify interesting patterns, etc.
Such operations require interactive visualizations (as opposed to a static image) in which graph elements are rendered as distinct visual objects; e.g., DOM objects in a web browser. 
This way, the user can manipulate the graph directly from the UI, e.g., click on a node or an edge to get additional information (metadata), highlight parts of the graph,  etc.
Given that graphs in many real-world scenarios are huge, the aforementioned visualizations pose significant technical challenges from a data management perspective.

First of all, the visualization must be feasible without the need to load the whole graph in main memory. These "holistic" approaches
\cite{BastianHJ09,HastrupCB08}
result in prohibitive memory requirements, and usually rely on dedicated client-server architectures which are not always affordable by enterprises, especially start-ups.
Then, the visualization tool must ensure extremely low response time, even in multi-user environments built upon commodity machines with limited computational resources.
Finally, the visualization must be flexible and meaningful to the user, allowing her to explore the graph in different ways and at multiple levels of detail.

State-of-the-art works in the field
\cite{AbelloHK06,ArchambaultMA08,Auber04,RodriguesTTFL06,LinCTWKC13,TominskiAS09,ZinsmaierBDS12}
tackle with the previous problems through a hierarchical visualization approach.
In a nutshell, hierarchical visualizations merge parts of the graph into abstract nodes (recursively) in order to create a tree-like structure of abstraction layers. This results in a decomposition of the graph into much smaller (nested) sub-graphs which can be separately visualized
and explored in a "vertical"  fashion, i.e., by clicking on an abstract node to retrieve the enclosed sub-graph of the lower layer.
In most cases, the hierarchy is constructed by exploiting clustering and partitioning methods \cite{AbelloHK06,Auber04,BastianHJ09,RodriguesTTFL06,TominskiAS09}.
In other works, the hierarchy is defined with hub-based \cite{LinCTWKC13} and density-based \cite{ZinsmaierBDS12} techniques. \cite{ArchambaultMA08} supports ad-hoc hierarchies which are manually defined by the users. 
A different approach has been adopted in \cite{SundaraAKDWCS10}
 where  sampling techniques are exploited. 
 Finally, in the context of the Web of Data, there is a large number of tools that visualize RDF graphs \cite{bs16}; however,
 all these tools require the whole graph to be loaded on the UI.
Although the hierarchical approaches provide fancy visualizations with low memory requirements, 
they do not support intuitive 
 "horizontal" exploration (e.g., for following paths in the graph).
Further,  with hierarchical approaches it is not easy to explore dense parts of the graph in full detail (i.e., without using an abstract representation). 
Finally, the applicability of hierarchical approaches is heavily based on the particular characteristics of the dataset; for example, the existence of small and coherent clusters
\cite{AbelloHK06,ArchambaultMA08,Auber04,RodriguesTTFL06,TominskiAS09} or the distribution of node degrees \cite{LinCTWKC13,ZinsmaierBDS12}.

\stitle{Contribution.}
We introduce a generic platform for scalable multi-level visual exploration of large graphs.
The proposed platform can easily support various visualizations, 
including all ad-hoc approaches in the literature, 
and bases its efficiency on a novel technique for indexing and storing the graph at multiple levels of abstraction.
In particular, our approach involves an offline preprocessing phase that builds the layout of the input graph by assigning coordinates to its nodes with respect to a Euclidean plane.
The same offline procedure is followed for all levels of abstraction. 
The respective points are then indexed with a spatial data structure (i.e., R-tree) and stored in a database.
This way, our system maps user operations into efficient spatial operations (i.e., window queries) in the backend.
The prototype we demonstrate here is a proof of concept that interactive visualizations can be effective on commodity hardware, still, allowing the user to perform intuitive navigations on the plane (e.g., follow  paths in the graph) 
at any level of abstraction and regardless the size of the graph.

 \begin{figure*}[t]
\centering
  \subfloat{\includegraphics[width=7.0in]{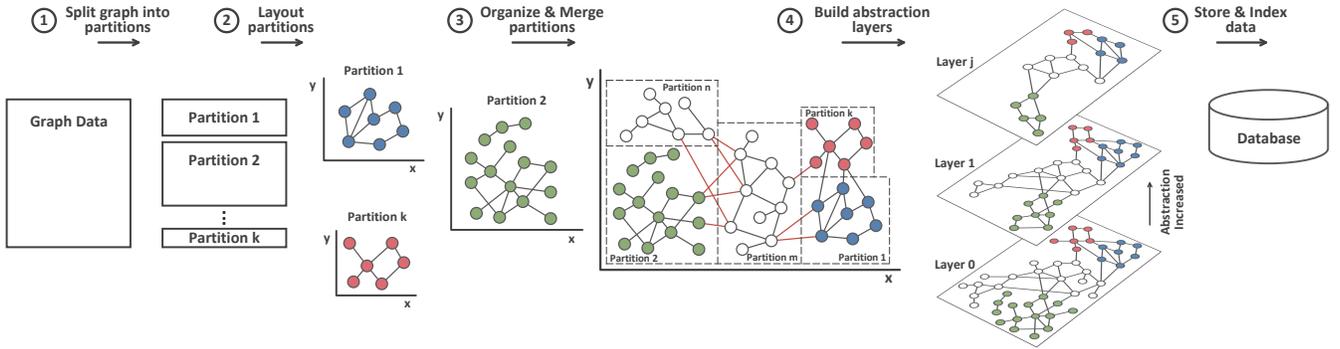}}
    \caption{Preprocessing Overview }
\label{fig:prep}
 \end{figure*}

\section{System Overview}
\label{sec:overview}

The architecture of our prototype, graphVizdb,
consists of three main parts: (1) the \emph{Client}, (2) the graphVizdb \emph{Core} module, and (3) the \emph{Database}.
The \emph{Client} is the frontend UI that offers
several functionalities to the users, e.g., an interactive canvas, search features, multi-level exploration, etc.
The \emph{Core} module contains the \emph{Preprocessing} module and the \emph{Query Manager} that is responsible for the communication between the \emph{Client} and the \emph{Database}.
The Preprocessing module contains the following submodules: \textit{Partitioning}, \textit{Layout}, \textit{Partition Organizer} and
\textit{Abstraction}.
Finally, the \emph{Database} contains all data needed for the visualization along with the necessary indexes.
Details for each part are provided in the following.



\subsection{Preprocessing}
\label{sec:prepro}

In our approach, the layout of the input graph is built on the server side once, during the preprocessing phase,
and this can be done with any of the existing layout algorithms.
The result of this process is the assignment of coordinates to the nodes of the graph with respect to a Euclidean plane.
The state-of-art layout algorithms provide layouts of high quality, however, they require large amounts of memory in practice, even for
graphs with few thousands of nodes and edges.
In order to overcome this problem, we adopt a partition-based approach as shown in Fig.~\ref{fig:prep}.

Initially, the graph is divided by the \emph{Partitioning} module into a set of $k$ distinct sub-graphs ($Step~1$), where $k$ is proportional to the total graph size and the available memory of the machine.
This is a $k$-$way$ partitioning that aims at minimizing the number of edges between the different sub-graphs \cite{KarypisK95}.
Then, the \emph{Layout} module applies the layout algorithm to each partition independently, and assigns coordinates to the nodes of each sub-graph without considering the edges that cross different partitions ($Step~2$). Any layout algorithm can be used in this step, e.g., circle, star, hierarchical, etc.
The edges between the different sub-graphs are taken into account by the \textit{Partition Organizer} when arranging the partitions on the "global" plane at $Step~3$. Multiple abstraction layers of the input graph are constructed by the \emph{Abstraction} module at $Step~4$. At $Step~5$, the input graph along with the abstract graphs are indexed and stored in the \emph{Database}.
In the following, we provide more details on $Step~3$, $4$, and $5$ of Fig.~\ref{fig:prep}.

\stitle{Organizing   Partitions.}
Partitions are organized on the "global" plane using a greedy algorithm
whose goal is twofold. First, it ensures that the distinct sub-graphs do not overlap on the plane, and at the same time it tries to minimize the total length of the edges between different partitions (crossing edges).

Initially, the algorithm counts the number of crossing edges for each partition. Then, it selects the partition with the largest number of crossing edges (to all other partitions), and places it at the center of the plane, i.e., it updates the coordinates of its nodes with respect to the "global" plane. This is the $m$-th partition in Fig.~\ref{fig:prep} which has $9$ such edges (denoted with red color). The remaining partitions are kept in a priority queue, sorted on the number of the common crossing edges they have with the partitions that exist on the plane (in descending order).
At each subsequent step, the algorithm assigns the first partition from the queue to an empty area on the plane so that the total length of the crossing edges between this partition and all other partitions on the plane is minimized. Then, the partition is removed from the queue and the coordinates of its nodes are updated with respect to the assigned area. The order of the partitions in the queue is also updated accordingly, and the algorithm proceeds to the next step.
The above process terminates when the priority queue is empty.
Intuitively, the efficiency of the algorithm is guaranteed by the small number of partitions ($k$), and also by the small size of the area we have to check for the best assignment at each step; this area lies around the non-empty areas from the previous steps.

\stitle{Building Abstraction Layers.}
After arranging the partitions,
a number of abstraction layers is constructed for the initial graph, as shown in Fig.~\ref{fig:prep}.
A layer $i$ ($i>0$) corresponds to a new graph that is produced by applying an abstraction method to the graph at layer $i-1$. Hence, the overall hierarchy of layers is constructed in a bottom-up fashion, starting from the initial graph at layer 0.
Each time we create a new graph at layer $i$, its layout is based on the layout of the graph at layer $i-1$.
The abstraction method can be any algorithm that produces a more condense form of the input graph, either by merging parts of the graph into single nodes (like the graph summarization methods we mentioned in the introduction) or by filtering parts of the graph according to a metric, e.g., a node ranking criterion like PageRank.
We emphasize that our approach does not pose any restrictions to the number of layers or the size of the graph at each layer.
Finally, all layers are kept as separate graphs in the database as we explain below.

\begin{figure}[h]
\centering
{
\includegraphics[trim = 0mm 0mm 0mm 0mm, width=2.675in]{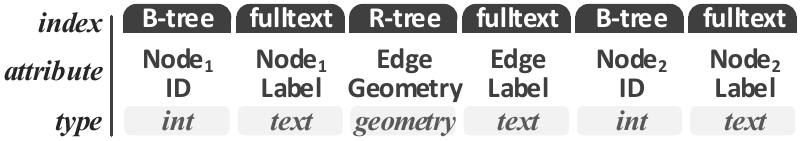}}
\caption{Storage Scheme}
\label{fig:scheme}
\vspace{-3mm}
\end{figure}

\stitle{Storage Scheme.}
Our database includes a single relational table per abstraction layer that stores all information about the graph of this layer.
All these tables have the same schema as depicted in Fig.~\ref{fig:scheme}.
Intuitively, each graph is stored as a set of triples of the form (node$_1$, edge, node$_2$).
A row in the table of Fig.~\ref{fig:scheme} contains the following attributes:
(1) the unique ID of the first node (Node$_1$ ID),
(2) the label of the first node (Node$_1$ Label),
(3) the geometry of the connecting edge (Edge Geometry) which is an binary object that represents the line between node$_1$ and node$_2$ on the plane,
(4) the label of the edge (Edge Label),
(5) the unique ID of the second node (Node$_2$ ID), and
(6) the label of the second node (Node$_2$ Label).
When the edge is directed, node$_1$ is always the source node whereas node$_2$ is the target node. This information is encoded in the binary object that represents the geometry of the edge.

B$^+$-trees are built on attributes (1) and (5) to retrieve all information about a node efficiently.
The full text indexes shown in Fig.~\ref{fig:scheme} correspond to tries, and they are used to support fast keyword search on the graph metadata.
Finally, an R-tree is used to index the geometries of the edges on the plane. Note that each such geometry is internally defined by the coordinates of the first and the second node whose IDs and labels are stored in the same row of the table.

\subsection{Online Operations}
\label{sec:operations}

On the client side, our system provides three main visual operations:

\stitle{Interactive Navigation.}
The user
navigates on the graph by moving the viewing window ("horizontal" navigation).
When the window is moved, its new coordinates with respect to the whole canvas are tracked on the client side, and a spatial range query (i.e., a window query) is sent to the server.
This query retrieves all elements of the graph (nodes and edges) that overlap with the current window. The query is evaluated with a lookup in the R-tree of Fig.~\ref{fig:scheme}, and the respective part of the graph is fetched from the database and sent to the client.
After the part of the graph is rendered on the canvas, the user can start the exploration. 


\stitle{Multi-level Exploration.}
The user moves up or down at different abstraction layers of the graph through a  \emph{Layer Panel} ("vertical" navigation). When changing a level of abstraction, the graph elements are fetched through spatial range queries on the appropriate table that corresponds to the selected layer. Vertical navigation can be combined with traditional zoom in/out operations in order to give the impression of a lower/higher perspective. In this case, the size of the window (rectangle) that is sent to the server is decreased/increased proportionally according to the zoom level.

\stitle{Keyword-based Exploration.}
Finally, the user searches the graph using keywords through a \emph{Search Panel}.
In this case, a keyword query is sent to the server and it is evaluated on the whole set of node labels which are indexed with tries. The result of this query is a list of nodes whose labels contain the given keyword.
By clicking on a node from the list, the user's window focuses on the position of this node. In this case, the spatial query sent to the server uses as window the rectangle whose size is equal to the size of the client's window and whose center has the same coordinates with the selected node from the list.



\section{Implementation \& Evaluation}
This section provides information on the implementation
of our prototype. It also includes the results of our experimental evaluation with two real graph datasets.

\stitle{Implementation.}
graphVizdb\footnote{\href{http://graphvizdb.imis.athena-innovation.gr}{{graphvizdb.imis.athena-innovation.gr}}} is implemented on top of several
open-source tools and libraries. The \emph{Core} module of our
system is developed in Java  1.7, and the database we use is MySQL 5.6.12.
The partitioning of the graph, during the preprocessing phase, is done with Metis 5.1.0
whereas the layout of each partition is built with Graphviz 2.38.0.
The web-based frontend is entirely based on HTML and JavaScript.
For the interactive visualization of the graph on the client side, we use \mbox{mxGraph} 3.1.2.1.

\stitle{Web UI.}
The user interface consists of the following panels:
(1)  \textit{Visualization}, i.e., the interactive canvas,
(2) \textit{Information} that provides information about
a selected node (metadata),
(3) \textit{Control} that offers the basic functionality (e.g., select dataset, zoom, abstraction level/criterion),
(4) \textit{Birdview}, i.e., a large-scale image of the whole graph on the plane,
(5) \textit{Search} that provides keyword search functionalities,
(6) \textit{Statistics} that offers basic statistics for the graph (e.g., average node degree, density, etc.),
(7) \textit{Filter} that provides   filtering operations on the canvas (i.e., hide edges/nodes), and
(8) \textit{Edit} that allows the user to store in the database the graph modifications made through the canvas.

\stitle{Performance Evaluation.}
The experiments we present here were conducted on the Okeanos cloud
using a VM with a quad-core CPU at 2GHz and 8GB of RAM running Linux.
For the client application, we used Google Chrome on a laptop with an i7 CPU at 1.8GHz and 4GB of RAM.
The cache size of MySQL on the server side was set to 6GB.


To evaluate the response time of our system, we used several real graph datasets with rather different characteristics.
Due to lack of space, here we present only the results for two datasets: the \mbox{\textit{Wikidata}}
 RDF dataset, and the \textit{Patent}
 citation graph.
The first one is an RDF export of Wikidata having $151$M edges and $146$M nodes. 
The second dataset is taken from the SNAP repository of large network datasets;
 it contains $16.5$M edges and $3.8$M nodes. 

Table~\ref{tab:prepr} presents the preprocessing time for each step of Fig. \ref{fig:prep}. These times are higher for Wikidata since it is much bigger than the Patent dataset. The only exception is the time spent in Step 1 for applying the \textit{k-way} partitioning; this process takes longer for Patent due to the higher average node degree. Note that the most expensive part of the preprocessing is the indexing step; however, the presented times correspond to the total time spent in indexing 5 layers of each dataset, one after the other. In practice, we can speed up this step by distributing the layers to different nodes of the cluster and perform the indexing in parallel.
In this case, the time spent in Step 5 equals the time for indexing the input graph (layer 0), that is, $274.5$ and $17.4$ minutes for Wikidata and Patent respectively.

\begin{figure}[]
\centering
\subfloat[{Wikidata}]{
\hspace{-6pt}
\includegraphics[   height=1.48in]{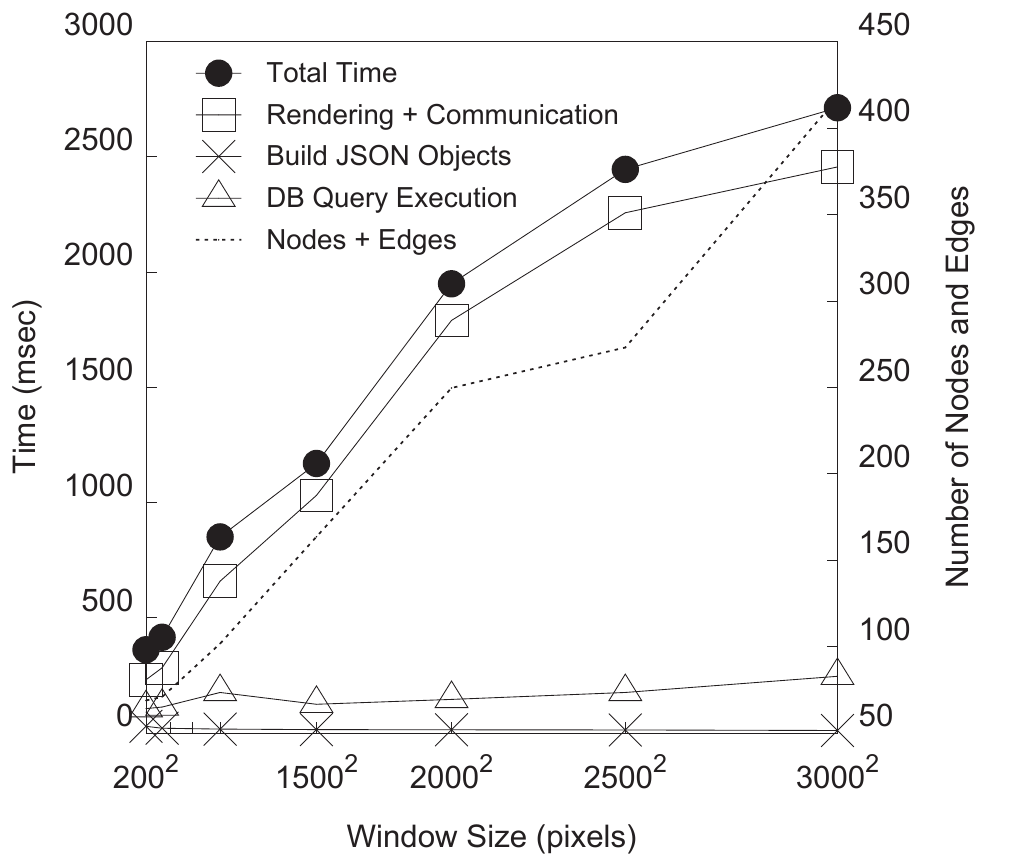} \label{fig:person_time} \hspace{-3mm}}
\subfloat[Patent] { \includegraphics[ height=1.48in]{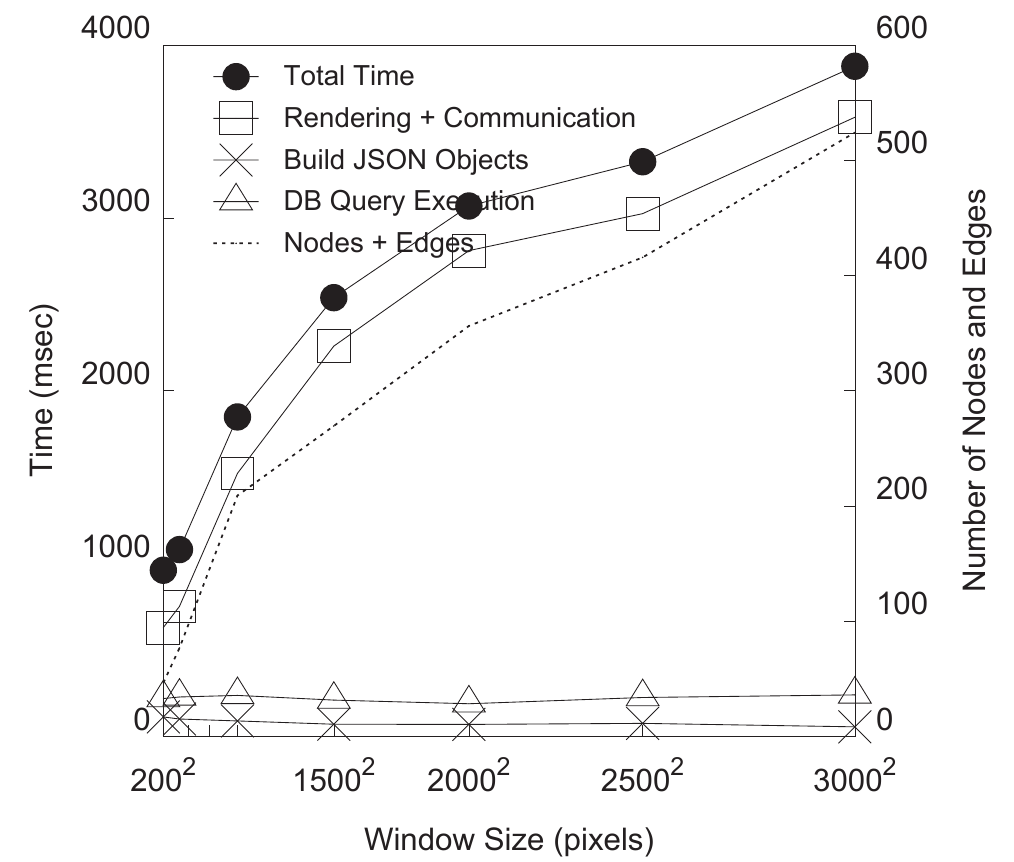}\label{fig:notre_time}}\\
\caption{Time vs.\ Window Size}
 \vspace*{-5mm}
\label{fig:results}
\end{figure}

Our experimental scenario includes the evaluation of window queries
with different sizes. These queries are evaluated by the server and sent to the client for visualization.
In particular, we used window queries whose size varies from $200^2$ to $3000^2$ pixels, and we evaluated
them on the initial graph of each dataset, i.e., on the bottom layer of abstraction (layer 0).
For each window size, we generated 100 random queries. The results we present in Fig.~\ref{fig:results} correspond to the following average times per query (msec):
(1) \textit{DB Query Execution}: the time spent to evaluate the query in the database,
(2) \textit{Build JSON Objects}: the time required for the server to process the query result and build the JSON objects that are sent to the client,
(3) \textit{Communication + Rendering}: the time spent in the client-sever communication plus the time needed to render the graph on the browser, and
(4) \textit{Total Time}: the sum of the above times.
The \textit{Nodes + Edges} in Fig. \ref{fig:results} refer to the average number of nodes and edges included in the 1K random windows of each size.

The first observation is that the performance of our approach scales linearly with the window size and the total number of objects in it. This behaviour is similar for both datasets.
As we can see in Fig.~\ref{fig:results}, the overall response time of the system is dominated by the time spent in \textit{Communication + Rendering}. We do not present these two operations separately because the part of the graph included in the window of the user is sent from the server to the client in small pieces, i.e., in a streaming fashion; hence, the respective times cannot be easily distinguished.
As a final comment, the time spent to evaluate the query in the database is negligible and increases slightly as the size of the window increases.

\begin{table}[]
\centering
\caption{Time for each Preprocessing Step (min)}
\label{tab:prepr}
\setlength{\tabcolsep}{4.8pt}
 \scriptsize
  \begin{tabular}{l||cc|ccccc}
\tline
\textbf{Dataset} &\textbf{\#Edges} &\textbf{\#Nodes} & \textbf{Step~1} &	\textbf{Step~2}  &	\textbf{Step 3} & \textbf{Step 4} &\textbf{Step 5}\\
 \dline
Wikidata  &151M& 146M & 1.8&	4.5 &	25.5 & 	16.5 & 670.1\\
Patent  & 16.5M & 3.8M &  5.1 &	2.8	& 9.7	& 8.2	& 41.2\\
\bline
\end{tabular}
\vspace{-1mm}
 \end{table}


\section{Demonstration Outline}

In this section, we outline our demonstration scenario.
%
%
The attendees will first select a dataset from a number of real-word datasets (e.g., ACM, DBLP, DBpedia).
Then, they will be able to have a quick glance on the graph using various navigation methods
such as panning, selecting a specific part of the graph in the birdview panel, etc.
Finally, the attendees will be able to filter (i.e., hide)  edges and/or nodes of specific types (e.g., RDF literals), as well as to zoom in/out over the graph.
For example, in the ACM dataset, a user interested in exploring the citations between articles
will be able to filter out irrelevant edges (e.g., \textit{has-author}, \textit{has-title}, etc.) and visualize only the \textit{cite} edges.

Additionally, the attendees will be able to explore the ``Focus on node'' mode,
which is suitable for pathway navigation, as well as for helping users to
further understand the relations amongst the nodes of interest.
In this mode, only the selected node and its neighbours are visible.
The user interested in exploring the scientific collaborations of an author will be
able to use keywords in order to search for this person, e.g., ``Christos Faloutsos''.
Then, using the ``Focus on node'', the user can quickly explore all Faloutsos' collaborations by following the
``Christos Faloutsos $\cdot$ \textit{has-author} $\cdot$ article $\cdot$ \textit{has-author}'' paths.

Beyond simple navigation, the attendees will be able to perform a multi-level graph exploration.
In particular, they will be able to modify the abstraction level as well as the abstraction criteria (i.e., Node degree, PageRank, HITS).
For example, by selecting either PageRank or HITS as the abstraction criterion in the Notre Dame web graph,
the users will be able to view different layers of the graph that contain only the ``important'' nodes (e.g., sites whose PageRank score is above a threshold).

A video presenting the basic functionality of our
prototype is available at:
\href{https://vimeo.com/117547871}{\myFontC{vimeo.com/117547871}}.



\stitle{Acknowledgement.} This work was partially supported by the EU project ``SlideWiki'' (688095).


\bibliographystyle{IEEEtran}
 
\bibliography{biblio}

\end{document}